\documentclass[peerreviewca, draftcls]{IEEEtran}
\IEEEoverridecommandlockouts 
\ifCLASSINFOpdf
\usepackage[pdftex]{graphicx}
\graphicspath{{./figure/ }}
\DeclareGraphicsExtensions{.pdf,.jpeg,.png}
\else
\usepackage[dvips]{graphicx}
\DeclareGraphicsExtensions{.eps}
\fi

\usepackage{tikz}
\usepackage{pgfplots}
\pgfplotsset{compat=1.15}
\usepackage[absolute,overlay]{textpos}
\usepackage[siunitx,european]{circuitikz}       
\usetikzlibrary{backgrounds}  
\usetikzlibrary{fit}          
\usetikzlibrary{positioning}  
\usetikzlibrary{chains}       
\usetikzlibrary{automata}     
\usetikzlibrary{trees}        
\usetikzlibrary{shadows}      
\usetikzlibrary{pgfplots.groupplots,calc,shapes.geometric,shapes.misc,arrows,matrix,fit,decorations.markings}
\tikzset{
  font={\fontsize{13.5pt}{16}\selectfont}}

\usepackage{subfigure}
\graphicspath{{figure/}}

\usepackage[utf8]{inputenc}
\usepackage[nolist]{acronym}
\usepackage[cmex10]{amsmath}
\usepackage{mdwmath}
\usepackage{mdwtab}
\usepackage{subfigure}
\graphicspath{{./figure/}}
\usepackage{afterpage}
\usepackage{multirow}
\usepackage{amsthm}
\usepackage{amssymb}

\usepackage{siunitx}
\DeclareSIUnit{\dBi}{dBi}
\DeclareSIUnit{\dBm}{dBm}
\DeclareSIUnit{\dBW}{dBW}
\usepackage{color}
\usepackage{dashrule}
\usepackage{algorithm} 
\usepackage{algpseudocode}
\usepackage{cite}
\usepackage{float}
\usepackage{svg}
\floatstyle{plaintop}
\usepackage{balance}

\usepackage{amssymb}

\newcommand{\C}{\ensuremath{\mathbb{C}}}

\begin{acronym}
	\acro{rx}[GS]{ground station}
	\acro{dl}[DL]{downlink}
	\acro{mimo}[MIMO]{multiple-input-multiple-output}
	\acro{ntn}[NTN]{non-terrestrial network}
	\acro{geo}[GEO]{geostationary orbit}
	\acro{meo}[MEO]{medium Earth orbit}
	\acro{leo}[LEO]{low Earth orbit}
	\acro{qos}[QoS]{quality of service}
	\acro{dip}[DiP]{distributed precoding}
	\acro{pspc}[PAPPC]{per-acces point power constraint}
	\acro{kkt}[KKT]{Karush-Kuhn-Tucker}
	\acro{svd}[SVD]{singular value decomposition}
	\acro{mse}[MSE]{mean-squared error}
	\acro{mmse}[MMSE]{minimum mean-squared error}
	\acro{zf}[ZF]{zero-forcing}
	\acro{mrt}[MRT]{maximum ratio transmission}
	\acro{mrc}[MRC]{maximum ratio combining}
	\acro{snr}[SNR]{signal-to-noise ratio}
	\acro{sinr}[SINR]{signal-to-interference-and-noise ratio}
	\acro{sic}[SIC]{successive interference cancellation}
	\acro{csi}[CSI]{channel state information}
	\acro{csit}[CSIT]{\ac{csi} at the transmitter}
	\acro{csir}[CSIR]{\ac{csi} at the receiver}
	\acro{los}[LOS]{line-of-sight}
	\acro{aod}[AoD]{angle of departure}
	\acro{aoa}[AoA]{angle of arrival}
	\acro{ula}[ULA]{uniform linear array}
	\acro{ecef}[ECEF]{Earth-centered, Earth-fixed }
	\acro{dft}[DFT]{discrete Fourier transform}
	\acro{sota}[SotA]{state of the art}
	\acro{rsma}[RSMA]{rate-splitting multiple access}
	\acro{sdma}[SDMA]{space-division multiple access}
	\acro{oma}[OMA]{orthogonal multiple access}
	\acro{noma}[NOMA]{non-orthogonal multiple access}
    \acro{tdma}[TDMA]{time-division multiple access}
    \acro{ofdma}[OFDMA]{orthogonal frequency-division multiple access}
	\acro{awgn}[AWGN]{additive white Gaussian noise}
	\acro{iui}[IUI]{inter-user interference}
	\acro{3gpp}[3GPP]{3rd Generation Partnership Project}
\end{acronym}

\allowdisplaybreaks

\usepackage{moreverb}
\immediate\write18{texcount -inc -incbib 
-sum main.tex > /tmp/wordcount.tex}

\newcommand{\txantennas}{N}
\newcommand{\numuser}{K}
\newcommand{\useridx}{k}
\newcommand{\txantenaidx}{n}
\newcommand{\usergain}{G_\mathrm{user}}
\newcommand{\satgain}{G_\mathrm{sat}}
\newcommand{\freq}{f}

\newcommand{\userkdist}{d_k}
\newcommand{\txantennadist}{d_N}
\newcommand{\steeringuserk}{\mathbf{a}_\useridx}
\newcommand{\wavelength}{\lambda}

\newcommand{\aodk}{{\theta}_k}
\newcommand{\eqaod}{{\theta}}
\newcommand{\error}{\varepsilon}

\newcommand{\receivesymbol}{y}

\newcommand{\precodingmatrix}{\mathbf{W}}

\newcommand{\noisevariance}{\sigma_n^2}
\newcommand{\signalvector}{\mathbf{x}}

\newcommand{\channelvector}{\mathbf{h}}
\newcommand{\channelvectork}{\mathbf{h}_k}
\newcommand{\precodingvectorcommon}{\mathbf{w}_c}
\newcommand{\precodingvectorprivate}{\mathbf{w}_{p,k}}
\newcommand{\precodingvector}{\mathbf{w}}

\newcommand{\sumrate}{R}
\newcommand{\commonpartidx}{c}
\newcommand{\privatepartidx}{p}
\newcommand{\txpower}{P}
\newcommand{\RSscalingfactor}{\alpha}
\newcommand{\exampleidx}{\eta_\useridx}
\newcommand{\Distanceuser}{D_\useridx}

\begin{document}	
\title{A Comparison between RSMA, SDMA, and OMA in Multibeam LEO Satellite Systems}

\author{
\IEEEauthorblockN{Alea Schröder\IEEEauthorrefmark{1},
Maik Röper\IEEEauthorrefmark{1}, Dirk Wübben\IEEEauthorrefmark{1}, Bho Matthiesen\textsuperscript{\IEEEauthorrefmark{1}\IEEEauthorrefmark{2}}, Petar Popovski\textsuperscript{\IEEEauthorrefmark{3},\IEEEauthorrefmark{2}} and Armin Dekorsy\IEEEauthorrefmark{1}}%
\IEEEauthorblockA{\IEEEauthorrefmark{1} Gauss-Olbers Center, c/o University of Bremen, Dept. of Communications Engineering, 28359 Bremen, Germany\\
	\IEEEauthorrefmark{2} University of Bremen, U Bremen Excellence Chair, Dept.\ of Communications Engineering, 28359 Bremen, Germany\\
\IEEEauthorrefmark{3} Aalborg University, Department of Electronic Systems, 9220 Aalborg, Denmark\\
	Email: \{schroeder,roeper,wuebben,matthiesen,dekorsy\}@ant.uni-bremen.de, petarp@es.aau.dk}
\thanks{This article is presented in part at the 2023 International ITG 26th Workshop on Smart Antennas and 13th Conference on Systems, Communications, and Coding.}
}

\maketitle

\begin{abstract}
    \Ac{leo} satellite systems enable close to global coverage and are therefore expected to become important pillars of future communication standards. However, a particular challenge faced by \ac{leo} satellites is the high orbital velocities due to which a precise channel estimation is difficult. We model this influence as an erroneous \ac{aod}, which corresponds to imperfect \ac{csit}. Poor \ac{csit} and non-orthogonal user channels degrade the performance of \ac{sdma} precoding by increasing \ac{iui}. In contrast to \ac{sdma}, there is no \ac{iui} in \ac{oma}, but it requires orthogonal time or frequency resources for each user.
    \Ac{rsma}, unifying \ac{sdma}, \ac{oma}, and \ac{noma}, has recently been proven to be a flexible approach for robust interference management considering imperfect \ac{csit}.
    In this paper, we investigate \ac{rsma} as a promising strategy to manage \ac{iui} in \ac{leo} satellite downlink systems caused by non-orthogonal user channels as well as imperfect \ac{csit}. 
    We evaluate the optimal configuration of \ac{rsma} depending on the geometrical constellation between the satellite and users.
    
    
\end{abstract}

\begin{IEEEkeywords}
	Low Earth orbit (LEO), Rate-Splitting Multiple Access, multi-user beamforming, MIMO satellite communications, beamspace MIMO, angle division multiple access, 3D networks
\end{IEEEkeywords}
\acresetall  

\section{Introduction}\label{sec:introduction}

Mobile networks are currently evolving from being focused on low-altitude and ground-based devices towards three-dimensional (3D) networks. Incorporating air- and spaceborne terminals into the sixth generation (6G) of mobile networks is expected to lead to ubiquitous global connectivity, a reduced carbon footprint of information and communication technology, and much higher resilience of the terrestrial network infrastructure. An integral step towards this goal is the integration of non-terrestrial networks (NTN) into the current infrastructure \cite{3GPP.TR.38.863, Leyva-Mayorga2020, Qu}.
%
In particular, \ac{leo} satellites have gained a lot of interest due to their significantly reduced latency, path losses, and deployment costs compared to conventional high-throughput satellites in \ac{meo} and \ac{geo} \cite{MaikBeamspace, Leyva-Mayorga2020}.
However, the low altitude of $\SI{500}-\SI{2000}{km}$ causes high velocities relative to terrestrial components. This is introducing a host of challenges including high Doppler shifts and imperfect channel state estimation due to imprecise position measurement.
These obstacles challenge conventional multiple access schemes like \ac{oma} and \ac{sdma}. The common \ac{oma} scheme since the fourth generation of mobile networks is to assign  orthogonal time or frequency resources for each user, i.e., a combination of \ac{ofdma} and \ac{tdma} \cite{Guidotti.etal.2019}. However, \ac{ofdma} is highly sensitive to Doppler shifts, while \ac{tdma} becomes very challenging for large and time-varying transmission delays \cite{Guidotti.etal.2019}.
On the other hand, \ac{sdma} enables higher spectral efficiencies compared to \ac{oma} but it requires multi-user precoding based on recent \ac{csi} in order to mitigate the mutual \ac{iui} \cite{Vandenameele}.
The effectiveness of the precoder, and correspondingly the mutual \ac{iui}, depends on the spatial separation of the users and the accuracy of the \ac{csit}.
High delays and relative velocities to the users on the ground impede the acquisition of \ac{csi} at the satellites.

A promising multiple access scheme in case of erroneous \ac{csit} is called \ac{rsma}.
While unifying \ac{sdma}, \ac{oma}, and \ac{noma} \cite{ClerckxUnifying}, \ac{rsma} has been shown to flexibly manage \ac{iui} for a wide range of applications \cite{ClerckxSurvey}. In \ac{rsma}, each user message is split into a private and a common part. The common part messages of all users are jointly encoded into a common data stream for all users. Depending on the amount of correlation between the user channels and errors of imperfections, different amounts of power are assigned to the common and private parts. Originally, \ac{rsma} has been investigated for use in terrestrial networks \cite{DaiMIMO} but has recently been extended to satellite use cases in \cite{YinSTIN, Vazquez2018}, among others. 
While investigation for terrestrial communications such as \cite{ChoiMUMIMO, DaiMIMO} tend to assume a Rayleigh fading channel, satellite communication is better covered by \ac{los} channel models \cite{Schwarz2019, MaikBeamspace}.
In \cite{ClerckxUnifying}, the authors investigated \ac{rsma} for different amounts of correlation between the user channels in a two-user scenario for \ac{los} and Rayleigh fading channels considering perfect \ac{csit}.
Our work complements their research by expanding their setting to a \ac{leo} satellite downlink scenario. We, therefore, consider a \ac{los} channel model and introduce an erroneous \ac{aod} to model imperfect position knowledge at the satellite. Furthermore, we study the optimal power allocation between common and private parts in \ac{rsma} depending on the level of correlation between the user channels while considering imperfect position knowledge. We compare the corresponding \ac{rsma} performance to \ac{oma} and \ac{sdma}. Our results show that \ac{rsma} outperforms both schemes by flexibly managing the \ac{iui}.

The paper is organized as follows. Section \ref{sec:system} presents the \ac{leo} downlink satellite system model as well as the conventional multiple access schemes, \ac{sdma} and \ac{oma}.
Section \ref{sec:RSMA} introduces the \ac{rsma} approach, which is numerically evaluated in section \ref{sec:simulation}. The results are concluded and discussed in Section \ref{sec:conclusion}.

\textit{Notation:} Lower and upper boldface letters denote vectors $\mathbf{x}$ and matrices $\mathbf{X}$, respectively. $\{\cdot\}^\text{T}$, $\{\cdot\}^\text{H}$ are indicating the transpose and complex conjugate transpose operator, while $\circ$ is the Hadamard product. The Euclidean norm is given by $\|\cdot\|$, absolute values by $|\cdot|$. 

\section{System Model}
\label{sec:system}

\subsection{\Ac{los} Channel}
\label{sec:los}
In this section, we introduce the \ac{leo} downlink \ac{los} system model as well as the assumed phase error model for imperfect \ac{csit}. One LEO satellite, equipped with a \ac{ula} containing $\txantennas$ antennas, serves $\numuser$ users. The users are handheld devices equipped with a single antenna with low receive gain $\usergain$. For the transmission from the satellite to user $\useridx$, we assume a \ac{los} channel vector $\channelvectork \in\C^{1\times N}$ and complex \ac{awgn} $n_\useridx \sim \mathcal{CN}(0,\noisevariance)$. The received signal $\receivesymbol_\useridx$ for a given signal vector $\signalvector$ and user $\useridx$ is described by

\begin{align}
\label{eq:transmission}
    \receivesymbol_\useridx = \sum_{\txantenaidx=1}^{\txantennas} \channelvector_{\useridx,\txantenaidx} \signalvector_\txantenaidx + n_\useridx,
\end{align}
where
$\signalvector\in\C^{N\times 1}$ is the linearly precoded signal vector, which differs depending on the chosen multiple access scheme and precoder design.

For \ac{leo} satellites the transmission is mainly characterized by the \ac{los} channel component.
A corresponding channel vector $\channelvectork $ for the $\useridx$-th user is given by \cite{You.Li.Wang.Gao.Xia.Ottersten.2020} 

\begin{align}
\label{eq:channel}
    \channelvectork( \cos(\aodk)) = 
    \sqrt{\usergain G_{\text{sat}}} \frac{\wavelength}{4\pi d_k}  \text{e}^{-j \varphi_\useridx}  \steeringuserk^\text{T} ( \cos(\aodk)),
\end{align}
where $\satgain$ is the satellite antenna gain, $\wavelength$ the wavelength and $\userkdist$ the distance from the satellite to user $\useridx$. The overall phase shift of the symbols from the satellite to user $\useridx$ is determined by $\varphi_\useridx \in [0,2\pi]$. The relative phase shifts between the $\txantennas$ satellite antennas towards user $\useridx$ are described by the steering vector $\steeringuserk(\cos(\aodk)) = [a^1_k(\cos(\aodk)), ... , a^N_k(\cos(\aodk))]^\text{T}$, where $\aodk$ is the \ac{aod} from the \ac{ula} origin to user $\useridx$. Fig. \ref{fig:geometry} shows the \acp{aod} $\aodk$ and $\eqaod_{\useridx-1}$ of two users $\useridx$ and $\useridx-1$ in a \ac{leo} satellite downlink scenario with a satellite altitude $d_0$ and inter-user distance $D_k$. In general, the $\txantenaidx$-th element of the steering vector $\steeringuserk{(\exampleidx)}$ for user $\useridx$ with a given argument $\exampleidx = \cos(\aodk)$ is given as

\begin{align}
\label{eq:steering}
     a_\useridx^\txantenaidx(\exampleidx) = \exp^{-j\pi \frac{\txantennadist}{\lambda} (\txantennas +1 - 2\txantenaidx) \exampleidx}
\end{align}
where $\txantennadist$ denotes the distance between the satellite antenna elements.

\begin{figure} [t]
	\begin{center}  
		\resizebox{0,57\linewidth}{!}{%
			\input{figures/verteilteSatellitenplus} 
		}
	\end{center}  
	\caption{Single satellite scenario with a satellite altitude $d_0$ and two users $k$ and $k-1$, positioned at the \acp{aod} $\aodk$ and $\eqaod_{k-1}$, characterized by the channel vectors $\channelvectork$ and $\channelvector_{\useridx-1}$ and their distance $\Distanceuser$.}
	\label{fig:geometry}  
\end{figure}

Due to the high velocity of \ac{leo} satellites, a precise estimation of the user positions might not be available at the satellite. 
The channel vector $\channelvectork( \cos(\aodk))$ of user $\useridx$ with the steering vector $\steeringuserk( \cos(\aodk))$ is highly dependent on the \ac{aod} $\aodk$ of user $\useridx$. In particular, the relative phase shifts between the satellite antenna elements, which determine the performance of \ac{sdma} precoding, are characterized by the \acp{aod} of the users \cite{RoeperTWC2022}. We, therefore, model imprecise position measurement as an erroneous \ac{aod}. Instead of adding an error directly on the \ac{aod} $\aodk$,  we consider an additive error on the $\cos(\aodk)$. This error is a uniformly distributed additive error $\error_\useridx\sim \mathcal{U}(-\Delta \error, + \Delta \error)$, that follows the same distribution for all user $\useridx$.
Therefore, and by using the definition of the steering vector \eqref{eq:steering}, we can interpret the additive phase error $\error_{\useridx}$ as an overall multiplicative error $\mathbf{a}_\useridx(\error_{\useridx})$ on the channel vector $\channelvectork$. Then, the estimated channel vector $\tilde{\mathbf{h}}_k$ can be written as

\begin{align}
\label{eq:error}
    \tilde{\mathbf{h}}_k\big( \cos(\aodk)\big) = \mathbf{h}_k\big( \cos(\aodk)\big) \circ \steeringuserk^\text{T}\big(\error_{\useridx}\big).
\end{align}
The precoding will be based on this channel estimation $\tilde{\mathbf{h}}_k$. In the following subsections, we introduce the multiple access schemes \ac{sdma} and \ac{oma}.

\subsection{\Acf{sdma}}

In \ac{sdma}, the user messages are spatially separated by beams that ideally steer a maximum amount of power into the user directions while minimizing the interference between the beams. Therefore, each user symbol $s_\useridx$ is weighted by a precoding vector $\precodingvector_{\useridx}$. Because the $\numuser$ users share the same time and frequency resources, the signal vector $\signalvector$ is given as a superposition of all weighted user symbols

\begin{align}
\label{eq:signalvectorSDMA}
    \begin{split}
        \signalvector &= \sum_{\useridx=1}^\numuser \precodingvector_\useridx s_k.
    \end{split}
\end{align}
We analyze the performance of the different precoding strategies by evaluating their corresponding sum rate $R$, which is determined by the \ac{sinr} of each user $\useridx$. Assuming Gaussian distributed transmit symbols, the sum rate is generally formulated as \cite{Tse2005},

\begin{align}
\label{eq:sumRate}
    \sumrate^\mathrm{SDMA} = \sum_{\useridx = 1}^{\numuser}  \log \left(1+\frac{\left|\channelvectork \precodingvector_{\useridx}\right|^2}{\noisevariance+ \sum_{i \neq k}^K|\channelvectork \precodingvector_{i}|^2}\right) .
\end{align}
In Section \ref{sec:RSMA} we extend this formulation to the \ac{rsma} method.

The \ac{mmse} precoder is a well-established and reliable \ac{sdma} precoder \cite{windpassinger2004detection, MMSEspace}. A corresponding precoding matrix $\precodingmatrix= [\precodingvector_{1} \dots \precodingvector_{\numuser}]$ is specified as

\begin{align}
	\label{eq:MMSE}
	\begin{split}
		&\mathbf{W}= \sqrt{\frac{P}{\text{tr} \{ {{\mathbf{W}^{\prime}}^{\text{H}}} \mathbf{W}^{\prime} \} }} \cdot \mathbf{W}^{\prime}  
\\[2ex]
	&\mathbf{W}^{\prime} = \Big[ \mathbf{\tilde{H}}^\text{H} \mathbf{\tilde{H}} + \sigma_n^2\cdot \frac{\numuser}{\txpower} \cdot \mathbf{I}_{N} \Big]^{-1} \mathbf{\tilde{H}}^\text{H} 
	\end{split},
\end{align}
where $\mathbf{\tilde{H}} = [\mathbf{\tilde{h}}_1 \dots \mathbf{\tilde{h}}_\numuser]^\text{T}$ is the complete estimated channel matrix according to \eqref{eq:error}. Due to this definition of the \ac{mmse} precoder, the transmit power $P$ is not always equally distributed among the $\numuser$ users. In particular, the transmit power of a user $\useridx$ with a bad channel, i.e., $\channelvectork$ has a small euclidean norm, is higher than for those with good channels. This behavior introduces some fairness among the different users but does not maximize the corresponding sum rate \eqref{eq:sumRate}.

\subsection{\Acf{oma}}

To not only evaluate the \ac{rsma} approach in comparison with \ac{sdma} precoding, we compare both approaches to an \ac{oma} approach. In \ac{oma}, all users receive their symbols via orthogonal time or frequency resources, such that there is no \ac{iui} on the received signals. 
In this case, the precoding approach maximizing the rate of user $\useridx$ is \ac{mrt}. It steers the maximum amount of the available transmit power per user into its direction. With equal power allocation among the users, this precoder is given by

\begin{align}
    \precodingvector_\useridx^\mathrm{MRT} = \sqrt{\frac{P}{K}} \cdot \frac{\tilde{\channelvectork}^\mathrm{H}}{\|\tilde{\channelvectork}\|}.
\end{align}
In contrast to the \ac{sdma} approach, each user is only assigned to $1/\numuser$th time or frequency resources, which limits the corresponding rate. Thus, the sum rate for \ac{oma} becomes

\begin{align}
\label{eq:oma}
    \sumrate^\mathrm{OMA} = \frac{1}{\numuser} \sum_{\useridx = 1}^{\numuser}  \log \left(1+\frac{\left|\channelvectork \precodingvector_{\useridx}\right|^2}{\noisevariance}\right) .
\end{align}
Note that the common \ac{oma} approaches \ac{ofdma} and \ac{tdma} are highly sensitive to the high Doppler shifts and delays of \ac{leo} satellite communication systems \cite{Guidotti.etal.2019}. In this paper, we neglect these influences on the transmission quality by assuming a perfect compensation of the effects. 

Next, we introduce the \ac{rsma} approach and define an achievable rate, similar to the sum rate, as the corresponding evaluation metric.

\section{\acl{rsma}}
\label{sec:RSMA}

\ac{rsma} flexibly manages \ac{iui} by partly decoding the interference with \ac{sic} and partly treating it as noise. To perform \ac{rsma} each user message $B_k$ is split into a private part $B_{\privatepartidx,k}$ and a common part $B_{\commonpartidx,k}$. Fig. \ref{fig:RS_Schematisch} presents a corresponding process flow diagram. In the following, the index $\privatepartidx$ marks private part entities, whereas $\commonpartidx$ indicates the common part. After splitting each user's message into a private and a common part, the common part messages $B_{c,k}$ of all $\numuser$ users are combined into one common part message $B_c$. The common part message $B_c$ and all private part messages $\{B_{p,k}\}_{k=1}^{\numuser}$ are encoded into symbols, such that each user is assigned one private symbol $s_k$ and the common symbol $s_c$. These symbols are precoded linearly and the resulting transmit vector $\signalvector\in\C^{N\times 1}$ follows as \cite{ClerckxSurvey},

\begin{align}
\label{eq:signalvector}
    \begin{split}
        \signalvector &= \precodingvectorcommon s_c +\sum_{\useridx=1}^\numuser \precodingvectorprivate s_k,
    \end{split}
\end{align}
where $\precodingvectorcommon \in\C^{N\times 1}$ is the common part precoding vector and $\precodingvectorprivate \in\C^{N\times 1}$ is the private part precoding vector of user $\useridx$. The design of these precoding vectors is pursued in the subsequent subsection, followed by the generalization of the sum rate for the \ac{rsma} case.

\begin{figure} [t]
	\begin{center}  
		\resizebox{0,8\linewidth}{!}{%
			\begin{tikzpicture}
	\node[draw,minimum height=0.5cm,minimum width=3cm,rotate=90,align=center,color={rgb:red,0;green,0.4470;yellow,0.7410}] at (4, -2.25) (d)  {Linear Precoder};
	
	\node[draw,minimum height=0.5cm,minimum width=3cm,rotate=90,align=center] at (2.75, -2.25) (enc)  {Encoder};
	
	\node[draw,minimum height=0.5cm,minimum width=1.75cm,rotate=90,align=center] at (1.35, -1.65) (mes)  {Message\\ Combiner};
	
	\node[draw,minimum height=0.3cm,minimum width=3cm,rotate=90,align=center] at (-0.4, -2.25) (splitter)  {Message Splitter};
	
	\node[draw,minimum height=0.5cm,minimum width=3cm,rotate=90,align=center] at (-2, -2.25) (data)  {User Data};

	\draw[decoration={markings,mark=at position 1 with
	{\arrow[scale=2,>=stealth]{>}}},postaction={decorate}]	(-1.75, -1)--(-0.75, -1) node[midway,below,yshift=0.025cm,xshift=-0.1cm] {\small$ {{B}}_{1}$} node[midway, below,yshift=-0.35cm] {$\vdots$};
    \draw[decoration={markings,mark=at position 1 with
	{\arrow[scale=2,>=stealth]{>}}},postaction={decorate}]	(-1.75, -2.25)--(-0.75, -2.25) node[midway,below,yshift=0.025cm,xshift=-0.1cm] {\small$ {{B}}_{k}$} node[midway, below,yshift=-0.35cm] {$\vdots$};
	\draw[decoration={markings,mark=at position 1 with
		{\arrow[scale=2,>=stealth]{>}}},postaction={decorate}]	(-1.75, -3.5)--(-0.75, -3.5) node[midway,below,yshift=0.025cm,xshift=-0.1cm] {\small$ {{B}}_{K}$};
	
	\draw[decoration={markings,mark=at position 1 with
		{\arrow[scale=2,>=stealth]{>}}},postaction={decorate}]	(-0.1, -1)--(0.8, -1) node[midway,below,yshift=0.025cm,xshift=0.1cm] {\small$ {{B}}_{c,1}$};
	\draw[decoration={markings,mark=at position 1 with
		{\arrow[scale=2,>=stealth]{>}}},postaction={decorate}]	(-0.1, -1.55)--(0.8, -1.55) node[midway,below,yshift=0.025cm,xshift=0.1cm] {\small$ {{B}}_{c,k}$} node[near start,above,yshift=-0.05cm,xshift=-0.09cm] {{\fontsize{2}{6} $\vdots$}};
	\draw[decoration={markings,mark=at position 1 with
		{\arrow[scale=2,>=stealth]{>}}},postaction={decorate}]	(-0.1, -2.1)--(0.8, -2.1) node[midway,below,yshift=0.025cm,xshift=0.1cm] {\small$ {{B}}_{c,K}$} node[near start,above,yshift=-0.05cm,xshift=-0.09cm] {{\fontsize{2}{6} $\vdots$}};
	
	    \draw[decoration={markings,mark=at position 1 with
		{\arrow[scale=2,>=stealth]{>}}},postaction={decorate}]	(1.9, -1.5)--(2.5, -1.5) node[midway,below,yshift=0.025cm] {\small$ {{B}}_c$};
		\draw[decoration={markings,mark=at position 1 with
		{\arrow[scale=2,>=stealth]{>}}},postaction={decorate}]	(3, -1.5)--(3.75, -1.5) node[midway,below,yshift=0.025cm] {\small$ {{s}}_c$};

		\draw[decoration={markings,mark=at position 1 with
		{\arrow[scale=2,>=stealth]{>}}},postaction={decorate}]	(-0.1, -2.7)--(2.5, -2.7) node[midway,below,yshift=0.025cm,xshift=0.05cm] {\small$ {{B}}_{p,1}$};
	\draw[decoration={markings,mark=at position 1 with
		{\arrow[scale=2,>=stealth]{>}}},postaction={decorate}]	(-0.1, -3.2)--(2.5, -3.2) node[midway,below,yshift=0.025cm,xshift=0.05cm] {\small$ {{B}}_{p,k}$} node[near start,above,yshift=-0.05cm,xshift=-0.5cm] {{\fontsize{2}{6} $\vdots$}};
	\draw[decoration={markings,mark=at position 1 with
		{\arrow[scale=2,>=stealth]{>}}},postaction={decorate}]	(-0.1, -3.7)--(2.5, -3.7) node[midway,below,yshift=0.025cm,xshift=0.05cm] {\small$ {{B}}_{p,K}$} node[near start,above,yshift=-0.05cm,xshift=-0.5cm] {{\fontsize{2}{6} $\vdots$}};

	\draw[decoration={markings,mark=at position 1 with
		{\arrow[scale=2,>=stealth]{>}}},postaction={decorate}]	(3, -2.7)--(3.74, -2.7) node[midway,below,yshift=0.025cm,xshift=0.05cm] {\small$ {{s}}_1$};
	\draw[decoration={markings,mark=at position 1 with
		{\arrow[scale=2,>=stealth]{>}}},postaction={decorate}]	(3, -3.2)--(3.74, -3.2) node[midway,below,yshift=0.025cm,xshift=0.05cm] {\small$ {{s}}_k$} node[near start,above,yshift=-0.05cm,xshift=-0.075cm] {{\fontsize{2}{6} $\vdots$}};
	\draw[decoration={markings,mark=at position 1 with
		{\arrow[scale=2,>=stealth]{>}}},postaction={decorate}]	(3, -3.7)--(3.74, -3.7) node[midway,below,yshift=0.025cm,xshift=0.05cm] {\small$ {{s}}_K$} node[near start,above,yshift=-0.05cm,xshift=-0.075cm] {{\fontsize{2}{6} $\vdots$}};
		
	\draw[decoration={markings,mark=at position 1 with
		{\arrow[scale=2,>=stealth]{>}}},postaction={decorate}] (4.25,-1) -- (5,-1)  node[midway, below] {${x}_1$} node[midway, below,yshift=-0.35cm] {$\vdots$};
	\draw[decoration={markings,mark=at position 1 with
		{\arrow[scale=2,>=stealth]{>}}},postaction={decorate}] (4.25,-2.25) -- (5,-2.25)  node[midway, below] {${x}_n$} node[midway, below,yshift=-0.35cm] {$\vdots$};
	\draw[decoration={markings,mark=at position 1 with
		{\arrow[scale=2,>=stealth]{>}}},postaction={decorate}] (4.25,-3.5) -- (5,-3.5)  node[midway, below] {${x}_N$};


\end{tikzpicture} 
		}
	\end{center}  
	\caption{Process Flow Diagram of \ac{rsma}. User data is split into a private and a common part. All common parts are combined into one message, which is destined for all users.}
	\label{fig:RS_Schematisch}  
\end{figure}

\subsection{Precoder Design}

An important design criterion in \ac{rsma} is the optimal allocation of total transmit power $\txpower$ to the precoding vectors of the private parts $\precodingvectorprivate $ and the common part precoding vector $\precodingvectorcommon  $. Therefore, a scaling factor $\RSscalingfactor \in [0,1]$ is introduced, such that the power allocation among the precoding vectors is given by

\begin{align}
\label{eq:power_allocation}
    \begin{split}
        & \|\precodingvectorcommon\|^2 = \txpower-\txpower^\RSscalingfactor\\
        &  \sum_{\useridx=1}^\numuser\| \precodingvector_{\privatepartidx,\useridx}\|^2 = \txpower^\RSscalingfactor.\\
    \end{split}
\end{align}
A scaling factor of $\RSscalingfactor = 1$ equals common \ac{sdma} precoding, whereas a theoretical factor of $\RSscalingfactor \rightarrow -\infty$ would correspond to multicast transmission.
In order to directly examine the influence of the scaling factor $\RSscalingfactor$ on the \ac{rsma} performance, we assume the same precoding methods for any given channel realization. 
For the common part precoder $\precodingvectorcommon $, we choose a precoder of the form \eqref{eq:commonpartprecoding}.

\begin{align}
\label{eq:commonpartprecoding}
    \precodingvectorcommon = \sqrt{\frac{P-P^\RSscalingfactor}{\txantennas}}\cdot [1 \dots 1]^\text{T}.
\end{align}
Though it is not optimized in regard to maximizing the sum rate, this precoder has the advantage of being independent of any \ac{csit}. For the private part we apply \ac{mmse} precoding according to \eqref{eq:MMSE}. In order to fulfill the power constraints from \eqref{eq:power_allocation}, the transmit power $\txpower$ is reduced to $\txpower^\alpha$, such that the private part precoding matrix $\mathbf{W}_\privatepartidx$ follows as

\begin{align}
	\label{eq:MMSE_RSMA}
	\begin{split}
		&\mathbf{W}_\privatepartidx= \sqrt{\frac{P^\RSscalingfactor}{\text{tr} \{ {{\mathbf{W}^{\prime}}^{\text{H}}} \mathbf{W}^{\prime} \} }} \cdot \mathbf{W}^{\prime}  
\\[2ex]
	&\mathbf{W}^{\prime} = \Big[ \mathbf{\tilde{H}}^\text{H} \mathbf{\tilde{H}} + \sigma_n^2\cdot \frac{\numuser}{\txpower^\RSscalingfactor} \cdot \mathbf{I}_{N} \Big]^{-1} \mathbf{\tilde{H}}^\text{H}
	\end{split}.
\end{align}

\subsection{Achievable Rate}

To evaluate the performance of \ac{rsma}, we need to extend the sum rate metric to the common part. With the signal vector $\signalvector\in\C^{N\times 1}$ from \eqref{eq:signalvector} and \eqref{eq:transmission}, the receive signal $y_k$ is given by

\begin{align}
	y_k =  \mathbf{h}_k\mathbf{w}_{c} s_c + \mathbf{h}_k{\mathbf{w}}_{p,k} s_k +{\mathbf{h}_k}\sum_{i\neq k}{\mathbf{w}}_{p,i} s_i  + {n_k}.
\end{align}
In \ac{rsma}, the common part symbol $s_c$ is the first to be decoded by \ac{sic}. This requires additional receiver complexity compared to \ac{sdma} precoding \cite{ClerckxSurvey}. For perfect \ac{sic}, the rate for the common part symbol $R_{c,k}$ of user $\useridx$ is given by

\begin{align}
    \sumrate_{\commonpartidx,\useridx} = \log \bigg(1+\frac{|\channelvectork \precodingvectorcommon|^2}{\noisevariance + \sum_{i =1}^K|\channelvectork \precodingvector_{p,i}|^2}\bigg),
\end{align}
%
that takes the interference of all private parts into account. The remaining received signal after the successful decoding of the common part is

\begin{align}
	y_{\privatepartidx, \useridx} =  \mathbf{h}_k{\mathbf{w}}_{p,k} s_k +{\mathbf{h}_k}\sum_{i\neq k}{\mathbf{w}}_{p,i} s_i  + {n_k}.
\end{align}
For the residual signal $y_{\privatepartidx, \useridx}$, the \ac{iui} will be treated as noise, such that the corresponding private rate $\sumrate_{\privatepartidx,\useridx} $ of user $\useridx$ results in

\begin{align}
    \sumrate_{\privatepartidx,\useridx} = \log \bigg(1+\frac{|\channelvectork \precodingvector_{\privatepartidx,\useridx}|^2}{\noisevariance + \sum_{i \neq k}^K|\channelvectork \precodingvector_{p,i}|^2}\bigg).
\end{align}
To get an equivalent rate compared to \eqref{eq:sumRate}, all private part rates are summed up, and the smallest common part rate is added to guarantee successful common part decoding at all users \cite{ClerckxSurvey}. 
The specific sum rate for \ac{rsma} precoding calculates as

\begin{align}
    \label{eq:sumrateRS}
	\begin{split}
	R^\mathrm{RSMA} &=  \min_k \sumrate_{\commonpartidx, \useridx}+ \sum_{k=1}^{K} R_{\privatepartidx,\useridx}.
	 \end{split}
\end{align}
%
Note that in this paper, we assume simple common part precoding \eqref{eq:commonpartprecoding} and \ac{mmse} private part precoding \eqref{eq:MMSE_RSMA}. We do not follow an optimal design based on \eqref{eq:sumrateRS}.

\section{Numerical Evaluations}
\label{sec:simulation}

In this paper, we investigate \ac{rsma} for a \ac{los} channel with multiplicative error \eqref{eq:error}. Given that \ac{rsma} was originally designed for Rayleigh fading channels with additive errors, our goal is to evaluate its suitability for robust interference management in \ac{leo} satellite downlink scenarios. 
All numerical evaluations in this section consider a two-user downlink scenario and the simulation parameters from Table \ref{tab:parameters}. This section is organized as follows. In Section \ref{sec:resultsperfect} we evaluate \ac{sdma} and \ac{rsma} for different realizations of the deterministic \ac{los} channel \eqref{eq:channel} assuming perfect \ac{csit}. In Section \ref{sec:resultsimperfect} we expand this evaluation to a scenario with imperfect \ac{csit}, e.g., an error on the \acp{aod} according to \eqref{eq:error}.

\begin{table}[!t]
    \renewcommand{\arraystretch}{1.3}
    \caption{Simulation Parameter}
    \label{tab:parameters}
    \centering
    \begin{tabular}{llc}
        \hline
        User Number & $\numuser$ & \SI{2}{}\\
        Satellite altitude & $d_0 $ & \SI{600}{km}\\
        Carrier frequency &$\freq$& \SI{2}{GHz}\\
        Satellite antenna number & $\txantennas$ & 6\\
        Inter-antenna-spacing & $\txantennadist$ & \SI{7.5}{cm}\\
        Satellite antenna gain & $\satgain$ & \SI{16}{dBi}\\
        User antenna gain & $\usergain$ & \SI{0}{dBi}\\
        Noise Power & $\noisevariance$ & \SI{-122}{dBW}\\
        Transmit Power & $P$ & \SI{20}{dBW}\\
        \hline
    \end{tabular}
\end{table}

\subsection{Perfect \ac{csit}}
\label{sec:resultsperfect}

The \ac{los} channel characteristics, as defined in Section \ref{sec:los}, highly differ depending on the given satellite altitude $d_0$, the number of transmit antennas $\txantennas$, their inter-antenna-distance $\txantennadist$, and the \acp{aod} of the users. In this paper, we solely focus on the impact of the \acp{aod} on the channel characteristics. The \ac{aod} $\eqaod_1$, see Fig. \ref{fig:geometry}, from the satellite to the first user is $90^\circ$. User $1$, therefore, is positioned directly underneath the satellite with the smallest satellite-to-user distance $d_1 = d_0 = \SI{600}{km}$. Whereas the position of user $2$, e.g., the \ac{aod} $\eqaod_2$, is variable. The difference $\Delta \eqaod$ between the \ac{aod} $\eqaod_2$ of user $2$ and the \ac{aod} $\eqaod_1$ of user $1$ is assigned to a corresponding user distance $\Distanceuser$. Depending on this difference $\Delta \eqaod$ and therefore depending on the user distance $\Distanceuser$, the characteristics of the \ac{los} channel vary. 

We introduce a correlation factor $\rho$ to quantify the amount of correlation between the channel vectors of both users,

\begin{align}
    \rho =  \frac{\channelvector^\text{H}_1 \channelvector_2}{\|\channelvector_1\| \|\channelvector_2\|}.
\end{align}
%
Accordingly, a correlation factor $\rho= 1$ corresponds to aligned channels, while a correlation factor $\rho = 0$ corresponds to orthogonal channels. Fig. \ref{fig:SVD} shows how the amount of correlation $\rho$ between the user channels decreases with an increasing user distance $D_k$ for $D_k\leq \SI{200}{km}$. 
\begin{figure} [t]
	\begin{center}  
		\resizebox{1\linewidth}{!}{%
			\input{figures/correlation.tex} 
		}
	\end{center}  
	\caption{Correlation factor $\rho$ depending on the User Distance $\Distanceuser$}
	\label{fig:SVD}  
\end{figure}

\begin{figure} [t]
	\begin{center}  
		\resizebox{1\linewidth}{!}{%
			\input{figures/WSA_plot.tex} 
		}
	\end{center}  
	\caption{Comparison of Achievable Rates between \ac{sdma}, \ac{oma} and different realizations of \ac{rsma} depending on the User Distance $\Distanceuser$ for perfect \ac{csit}. $\mathrm{RSMA}$, $\mathrm{opt.}$ corresponds to the optimal scaling factor $\RSscalingfactor_\mathrm{opt}$ for any given User Distance $\Distanceuser$.}
	\label{fig:w}  
\end{figure}

Fig. 4 depicts the achievable rates of \ac{sdma}, \ac{oma}, and \ac{rsma} for user distances $D_k$ between \SI{0.5}{km} and \SI{200}{km}. \ac{sdma} precoding performs best for orthogonal user channels. The higher the correlation factor $\rho$, the higher the \ac{iui}, which cannot be compensated by common \ac{sdma} precoding. To manage this mutual \ac{iui}, we apply \ac{rsma}. The \ac{rsma} approach is shown for different power scaling factors $\RSscalingfactor$. The larger $\RSscalingfactor$ the more power is assigned to the private parts, i.e. a scaling factor $\RSscalingfactor =1$ equals \ac{sdma} precoding. \Ac{rsma} opt. refers to an optimal $\RSscalingfactor_{\mathrm{opt}}$ for each user distance $D_k$ and is obtained by exhaustive search.  \ac{rsma} opt. outperforms the \ac{oma} approach for all values of $\Distanceuser$. Further, it is showing strong performance gains compared to \ac{sdma} for user distances $D_k$ up to roughly \SI{80}{km}. For larger user distances \ac{rsma} opt. and \ac{sdma} attain the same achievable rate. Therefore, the optimal $\RSscalingfactor$ for $D_k> \SI{80}{km}$ is $\RSscalingfactor_{\mathrm{opt}}|_{D_k> \SI{80}{km}}=1$. For user distances $\Distanceuser$ smaller than \SI{48}{km}, which corresponds to the intersection between \ac{sdma} and \ac{oma}, \ac{rsma} opt. follows the line of \ac{rsma} for $\RSscalingfactor = 0$. It can be observed that the performance of \ac{rsma} for $\RSscalingfactor= 0$ and therefore the performance of \ac{rsma} opt. slightly decreases in this area with increasing user distance $\Distanceuser$. This is due to the likewise decreasing correlation between the user channels, which is favorable for the transmission of the private parts but disadvantageous for the transmission of the common part, which is intended for all users. For $\RSscalingfactor= 0$ and an overall transmit power of $\txpower = \SI{20}{dBW}$ \SI{99}{\%} of the transmit power is assigned to the common part. Whereas there is no transmit power in the common part for $\RSscalingfactor=1$. Only for user distances $\Distanceuser$ between \SI{48}{km} and \SI{80}{km} scaling factors other than zero or one achieve the optimal \ac{rsma} performance for the given precoder designs. In general, it can be concluded that the higher the correlation between the channels, the smaller the optimal scaling factor $\RSscalingfactor_\mathrm{opt}$ and vice versa.

\subsection{Imperfect \ac{csit}}
\label{sec:resultsimperfect}

\begin{figure} [t]
	\begin{center}  
		\resizebox{1\linewidth}{!}{%
			\input{figures/WSA_plot_error.tex} 
		}
	\end{center}  
	\caption{Comparison of Achievable Rates between \ac{sdma}, \ac{oma} and different realizations of \ac{rsma} depending on the User Distance $\Distanceuser$ for imperfect \ac{csit}, e.g., an uniformly distributed additive error $\error = 0.2$ on the cosine of the \acp{aod} of both users. $\mathrm{RSMA}$, $\mathrm{opt.}$ corresponds to the optimal scaling factor $\RSscalingfactor_\mathrm{opt}$ for any given User Distance $\Distanceuser$.}
	\label{fig:h}  
\end{figure}
To expand the evaluation of \ac{rsma} in \ac{leo} satellite downlink scenarios, we now consider a scenario with imperfect \ac{csit}. Therefore, a uniformly distributed error $\error_\useridx$ with $\Delta \error = 0.2$ (see Section \ref{sec:los}) is applied on the cosine of the \acp{aod} of the users. It can be interpreted as uncertainty in the user positions. Fig. 5 shows the corresponding achievable rates for \ac{sdma}, \ac{oma}, and \ac{rsma} averaged over \SI{10000} Monte Carlo iterations. As expected the performance of common \ac{sdma} precoding is reduced compared to the scenario with perfect \ac{csit}. The maximum achievable rate, for example, dropped from \SI{6.1}{bps/Hz} to \SI{4.5}{bps/Hz} for a user distance of $\Distanceuser =  \SI{200}{km}$. Further, in the perfect \ac{csit} scenario \ac{sdma} outperformed \ac{oma} for user distances $\Distanceuser$ larger than roughly \SI{48}{km}, whereas the intersection point for the imperfect \ac{csit} scenario is at approximately \SI{58}{km}. Note that the achievable rate for \ac{oma} is also decreased by \SI{13}{\%} to \SI{2.7}{bps/Hz} compared to the perfect \ac{csit} scenario. In contrast to that, \ac{rsma} opt. is able to maintain a rate of at least \SI{3.8}{bps/Hz} over the complete interval of $\Distanceuser$. Though the maximum achievable rate for $\Distanceuser = \SI{200}{km}$ of the perfect \ac{csit} case is not obtained, \ac{rsma} opt. still outperforms common \ac{sdma} for user distances up to \SI{120}{km} compared to \SI{80}{km} in the perfect \ac{csit} scenario. The area in which \ac{rsma} prevails over common \ac{sdma} precoding is therefore increased by imperfect \ac{csit}.

\section{Conclusion and Discussion}\label{sec:conclusion}

In this paper, \ac{rsma} was applied in a \ac{leo} satellite downlink scenario with a \ac{los} channel and a multiplicative error on the channel vector. For the precoding of the common and the private parts, we implemented well-known and feasible precoding strategies. It was shown that \ac{rsma} is able to flexibly manage \ac{iui} for correlated user channels and imperfect \ac{csit} by adapting its power scaling factor to the given channel realization. It thereby prevails common \ac{sdma} precoding. In practice, of course, the optimal scaling factor can not be found by exhaustive search for every channel realization. Therefore, further study is required on how to utilize \ac{rsma} in satellite communication scenarios.

\section*{Acknowledgment}
This research was supported in part by the German Federal Ministry of Education and Research (BMBF) within the projects Open6GHub under grant number 16KISK016 and 6G-TakeOff under grant number 16KISK068 as well as by the German Research Foundation (DFG) under Germany's Excellence Strategy (EXC 2077 at University of Bremen, University Allowance).

\bibliographystyle{./my_lib/IEEEtran}
\bibliography{./my_lib/IEEEabrv,./my_lib/references}

\end{document}